\documentclass[onecolumn,12pt]{article}

\usepackage{graphicx} 
\usepackage{setspace} 
\usepackage{graphics,epsfig}
\usepackage{amsmath}
\usepackage{amssymb}
\usepackage{bm}
\usepackage{xcolor}

\usepackage{subfigure}
\usepackage{txfonts}

\usepackage{multirow}

\newcommand*{\obs}{\ensuremath{\mathbf{y}}}

\UseRawInputEncoding

 \usepackage[top=2cm, bottom=4cm, left=2.25cm, right=2.25cm]{geometry}

\begin{document}

\title{Automatic tempered posterior distributions for Bayesian inversion problems}
\author{L. Martino$^{\dagger}$, F. Llorente$^{*}$, E. Curbelo$^{*}$, J. L{\'o}pez-Santiago$^{*}$, J. M{\'i}guez$^{*}$\\
{\small $^{\dagger}$ Universidad rey Juan Carlos (URJC), Madrid, Spain. }\\
{\small  $^{*}$ Universidad Carlos III de Madrd (UC3M),  Madrid, Spain.} \\
}

\maketitle

\begin{abstract}
We propose a novel adaptive importance sampling scheme for Bayesian inversion problems where the inference of the variables of interest and the power of the data noise is split. More specifically, we consider a Bayesian analysis for the variables of interest (i.e., the parameters of the model to invert), whereas we employ a maximum likelihood approach for the estimation of the noise power. The whole technique is implemented by means of an iterative procedure, alternating sampling and optimization steps.  Moreover, the noise power is also used as a tempered parameter for the posterior distribution of the the variables of interest. Therefore, a sequence of tempered posterior densities is generated, where the tempered parameter is automatically selected according to the actual estimation of the noise power.  A complete Bayesian study over the model parameters and the scale parameter can be also performed.
 Numerical experiments show the benefits of the proposed approach.
\end{abstract}

\maketitle

\section{Introduction}

 The estimation of unknown parameters from noisy observations is an essential problem in signal processing, statistics and machine learning \cite{Fitzgerald01,Andrieu:MCMachineLearning2003,MartinoSigPro10}. Within the Bayesian signal processing framework, these problems are addressed by constructing posterior probability distributions of the unknowns. Given the posterior, one often wants to make inference about the unknowns, e.g., if we are estimating parameters, finding
the values that maximize their posterior, or the values that minimize some cost function given the uncertainty of the parameters. Unfortunately, obtaining closed-form solutions, usually expressed as integrals of the posterior, is infeasible in most practical applications. Hence, developing approximate computational techniques (such as importance sampling and {Markov chain Monte Carlo} (MCMC) algorithms) are often required \cite{Robert04,Liu04b,MARTINO_book}.
\newline
\newline
 The so-called {\it tempering of the posterior} is a well-known procedure for improving the performance of the Monte Carlo (MC) algorithms \cite{kirkpatrick83,Marinari92,Friel08,Moral06}. The tempering is obtained by modulating an artificial scale parameter or by sequentially including new data. The reasons of the improvement in the performance are several: improving mixing, discovering modes, foster the exploration of the inference space, etc. In the first iterations of the MC scheme, a posterior density with a bigger scale is considered. The artificial scale parameter (often called ``temperature'') is reduced during the iterations, until considering the true posterior distribution. However, the user should decide a {\it temperature schedule}, i.e., a decreasing rule for the scale parameter, which is usually chosen in an heuristic way. In the literature, the tempering procedure has gained a particular attention for the estimation of the marginal likelihood (a.k.a., Bayesian model evidence) \cite{Friel08,Neal01,Llorente20}. 
\newline
\newline
Furthermore, the joint inference of parameters (denoted as $\boldsymbol{\theta}$) of observation models, ${\bf f}(\boldsymbol{\theta})$, and scale parameters  of the likelihood function (that, in the scalar case, is usually denoted as $\sigma$) can be a hard task. Indeed, ``wrong choices'' of $\sigma$ values can easily jeopardize the sampling of ${\bm \theta}$.  In this work, we introduce a procedure to tackle this problem.
\newline
 \newline
More specifically, in this work, we design an adaptive importance sampling (AIS) scheme \cite{Bugallo15} for Bayesian inversion problems, where an automatic tempering procedure is implemented. We consider that  the vector of observations ${\bf y}$ is obtained by a nonlinear transformation ${\bf f}(\boldsymbol{\theta})$ of the variables of interest $\boldsymbol{\theta}$, perturbed by additive Gaussian noise with unknown power $\sigma^2$. The nonlinear mapping ${\bf f}(\boldsymbol{\theta})$ usually represents a complex physical model or a computer code etc. The resulting posterior densities are usually highly multimodal and complex distributions.
Furthermore, the inference task in the joint space $[\boldsymbol{\theta},\sigma ]$ is particularly challenging.   We proposed a split strategy to tackle this problem. We consider an optimization approach over $\sigma$ and a sampling scheme for ${\bm \theta}$. More specifically, we design an iterative procedure where these two tasks are alternated. Additionally, the actual maximum likelihood (ML) estimation of the noise power, $\widehat{\sigma}_{\texttt{ML}}^2$, is employed as a tempering parameter, starting from high values and then ``cooling down'' according to the ML estimations at each iteration. Therefore, the proposed scheme deals with a sequence of tempered posteriors according to the current estimation $\widehat{\sigma}_{\texttt{ML}}^2$.
It is important to observe that, given a fixed vector $\boldsymbol{\theta}$, the ML estimation $\widehat{\sigma}_{\texttt{ML}}^2$ can be obtained analytically.
 \newline
 Furthermore, the complete Bayesian analysis regarding the joint posterior of ${\bm \theta}$ and $\sigma$ is also possible (as discussed in Section \ref{SectSuperFer}). This is obtained by implementing a proper re-weighting of the samples generated by { the proposed algorithm, called Automatic Tempering AIS (ATAIS)}, without any additional evaluations of the observation model. 
  An approximation of the marginal posterior of $\sigma$ is provided as well. 
 The advantages of the proposed scheme are shown in two numerical experiments, one of them considering a complex astronomical model.


\section{Problem Statement}
\label{sec:bayes}

%

Let us denote the observed measurements as ${\bf y}=[y_1,...,y_K]^{\top}\in \mathbb{R}^{K}$, and the variable of interest  that we desire to infer, as ${\bm \theta}=[ \theta_1,..., \theta_M]^{\top} \in {\bm \Theta} \subseteq \mathbb{R}^{M}$.
Furthermore, let  us consider the observation model 
\begin{equation}
\mathbf{y} = {\bf f}(\boldsymbol{\theta}) + \mathbf{v},
\label{eq:model}
\end{equation}
where we have a nonlinear mapping,
\begin{eqnarray}
{\bf f}(\boldsymbol{\theta})=[f_1({\bm \theta}),...,f_K({\bm \theta})]^{\top}: {\bm \Theta} \subseteq \mathbb{R}^{M} \rightarrow   \mathbb{R}^{K}, 
\end{eqnarray}
and a Gaussian perturbation noise,
\begin{eqnarray}
\mathbf{v}=[v_1,...,v_K]^{\top}Ê\sim \mathcal{N}({\bf v}|{\bf 0}, \sigma^2 {\bf I}_{K}),
\end{eqnarray}
with $\sigma>0$, and we have denoted the $K$-dimensional unit matrix as ${\bf I}_{K}$.{\footnote{{ The model can be easily extended to a matrix of observations ${\bf Y}=[{\bf y}_1,...,{\bf y}_K]^{\top}\in \mathbb{R}^{d_y\times K}$ instead of a vector, if the  nonlinear mapping is of type ${\bf F}(\boldsymbol{\theta})=[{\bf f}_1({\bm \theta}),...,{\bf f}_K({\bm \theta})]^{\top}: {\bm \Theta} \subseteq \mathbb{R}^{M} \rightarrow   \mathbb{R}^{d_y \times K} $.   }}} The noise variance $\sigma^2$ is unknown, in general.  The mapping ${\bf f}(\boldsymbol{\theta})$ could be analytically unknown:  the only assumption is that we are able to evaluate it pointwise. The likelihood  function is 
\begin{eqnarray}
 \ell({\bf y}|\boldsymbol{\theta},\sigma)&=&\frac{1}{(2\pi\sigma^2)^{K/2}}\exp\left( -\frac{1}{2\sigma^2} ||\mathbf{y} - {\bf f}(\boldsymbol{\theta}) ||^2\right), \\
 &=&\frac{1}{(2\pi\sigma^2)^{K/2}}\exp\left( -\frac{1}{2\sigma^2} \sum_{k=1}^K (y_k-f_k({\bm \theta}))^2\right).
\label{eq:LH}
\end{eqnarray}
Note that we have two types of variables of interest: the vector $\boldsymbol{\theta}$ contains the parameters of the nonlinear mapping ${\bf f}(\boldsymbol{\theta})$, whereas $\sigma$ is a scale parameter of the likelihood function.
\newline 
Given the vector of measurements ${\bf y}$, we desire to make inferences regarding the hidden parameters ${\bm \theta}$ and the noise power $\sigma^2$, obtaining at least some point estimators ${\widehat {\bm \theta}}$ and ${\widehat \sigma}^2$. 
We are also interested in performing uncertainty and correlation analysis among the components of $\bm{\theta}$.
Furthermore, we aim to perform model selection, i.e., to compare, select  or properly average different models.
\newline
\newline
{\bf Bayesian inference in the complete space.}  We consider prior densities $g_\theta({\bm \theta})$ and $g_\sigma(\sigma)$ over the unknowns. 
  Hence, the complete posterior density is
\begin{eqnarray}
\label{Total_post}
p({\bm \theta},\sigma|{\bf y})= \frac{1}{p({\bf y})} p({\bm \theta},\sigma,{\bf y})= \frac{1}{p({\bf y})}\ell({\bf y}|\boldsymbol{\theta},\sigma) g_\theta({\bm \theta})g_\sigma(\sigma), 
\end{eqnarray}
The marginal likelihood $Z=p({\bf y})$ is 
\begin{eqnarray}
\label{MargLike_total}
Z=p({\bf y})= \int_{\mathbb{R}^+} \int_{{\bm \Theta}} \ell({\bf y}|\boldsymbol{\theta},\sigma) g_\theta({\bm \theta})g_\sigma(\sigma) d{\bm \theta} d\sigma,
\end{eqnarray}
This quantity is often needed for model selection. Since $Z({\bf y})$ is generally unknown, we  can usually evaluate pointwise the unnormalized posterior $\pi({\bm \theta},\sigma|{\bf y}) = \ell({\bf y}|\boldsymbol{\theta},\sigma) g_\theta({\bm \theta})g_\sigma(\sigma)$ (i.e., $p({\bm \theta},\sigma|{\bf y})\propto\pi({\bm \theta},\sigma|{\bf y})$).    
 More generally, the computation of integrals of the form
\begin{equation}\label{INT_gen}
 I= \int_{\mathbb{R}^+} \int_{{\bm \Theta}} h({\bm \theta},\sigma)  p({\bm \theta},\sigma|{\bf y}) d{\bm \theta} d\sigma,
\end{equation}
where $h(\cdot): {\bm \Theta}\times \mathbb{R}^+\rightarrow   \mathbb{R}$ is an integrable function, is usually required. We consider a Monte Carlo quadrature approach for approximating the integral above and, more generally, provide a particle approximation of the joint posterior $p({\bm \theta},\sigma|{\bf y})$.
\newline
\newline
{\bf Main observation.} Generally, generating random samples from a complicated posterior in Eq. \eqref{Total_post} and computing efficiently the integrals as in Eqs. \eqref{MargLike_total}-\eqref{INT_gen} is a hard task. Moreover, this task becomes often more difficult when we try to perform a joint inference where  scale parameters are involved, i.e., $\sigma$, and  parameters of the nonlinearity, i.e.,  ${\bm \theta}$. Indeed, ``wrong choices'' of $\sigma$ values can easily jeopardize the sampling of ${\bm \theta}$.  In the next section, we describe a strategy that we propose to tackle this problem. Before, we need to recall some additional definitions.
\newline
\newline
 {\bf Conditional and marginal posteriors.} In other to design efficient computational schemes
is often useful to consider the conditional posteriors, for instance, 
\begin{align}
p({\bm \theta}|{\bf y},\sigma)  = \frac{p({\bm \theta},{\bf y},\sigma)}{p({\bf y},\sigma)}& =\frac{\ell({\bf y}| {\bm \theta}, \sigma)g_\theta({\bm \theta})g_\sigma(\sigma)}{p({\bf y}| \sigma)g_\sigma(\sigma)}, \nonumber
\\ 
&
 = \frac{\ell({\bf y}|\sigma,{\bm \theta}) g_\theta({\bm \theta})}{p({\bf y}|\sigma)}.\label{EstaEslaConDMargPost}
\end{align}
In the next section, we will see that the idea underlying the proposed scheme is to split the space $[{\bm \theta},\sigma]$, restricting the sampling problem only with respect to ${\bm \theta}$ and considering an optimization problem with respect to $\sigma$.  The conditional marginal likelihood is obtained by integrating out one of the two variables, i.e.,
\begin{gather}
    \begin{split}
Z(\sigma)=p({\bf y}|\sigma)=\int_{{\bm \theta}} \ell({\bf y}|{\bm \theta},\sigma) g_\theta({\bm \theta})d{\bm \theta}.  
    \end{split}
    \label{condmargLike}
\end{gather}
 The integral above cannot be computed analytically, in general.
We can also consider marginal posteriors, for instance, the  marginal posterior of $\sigma$ is
\begin{align}\label{EqMargPost}
p(\sigma|{\bf y})=\frac{p({\bf y}|\sigma)g_\sigma(\sigma)}{p({\bf y})}=\frac{Z(\sigma)g_\sigma(\sigma)}{Z}.
\end{align}
Note that the joint posterior in Eq. \eqref{Total_post} can be also written as 
\begin{align}
\label{SuperIMPEq}
p({\bm \theta},\sigma|{\bf y}) &=p({\bm \theta}|{\bf y},\sigma)p(\sigma|{\bf y}).
\end{align}
{\bf Underlying idea.} The underlying idea of this work is to divide the inference study in two parts.
In the first part (Sections \ref{Sect_KeyFer} and  \ref{ATAIS_sect}), we focus on the study of the conditional posterior $p({\bm \theta}|{\bf y},\sigma)$ given a fixed $\sigma$. Then,  in the second part (Section \ref{SectSuperFer}),  we also estimate the marginal posterior $p(\sigma|{\bf y})$.
Finally, using \eqref{SuperIMPEq}, we can obtain a final approximation of the complete posterior $p({\bm \theta},\sigma|{\bf y})$. Estimations of $Z(\sigma)$ and $Z$ are also obtained.
   
\section{Key observations and proposed approach} \label{Sect_KeyFer}

In the first part of work,  we assume a proper (or improper) uniform prior over ${\bm \theta}$, i.e., $g_{\theta}(\bm{\theta})\propto 1$ in ${\bm \Theta}$. The possible use of a general choice of $g_{\theta}(\bm{\theta})$ is discussed in Section  \ref{GenPriorG}.  Let $\bm{\theta}_\texttt{MAP}$ denote the MAP estimator of $p(\bm{\theta}|\obs,\sigma)$. Generally, $\bm{\theta}_\texttt{MAP}$ should be a function of $\sigma$, i.e., $\bm{\theta}_\texttt{MAP}=\bm{\theta}_\texttt{MAP}(\sigma)$. However,  due to the choice of likelihood function (and the uniform prior) considered in this paper, we have that 
\begin{align*}
	\bm{\theta}_\texttt{MAP} 
	 &=\arg\max_{\bm{\theta}} \log p(\bm{\theta}|\obs,\sigma) \qquad  \mbox{ with }  \quad g_{\theta}({\bm \theta})\propto 1, \mbox{ for }  {\bm \theta} \in{\bm \Theta},\\
	 &=\arg\min_{\bm{\theta}} ||\obs-{\bf f}(\bm{\theta})||^2,
\end{align*}
 which  does not depend on $\sigma$, i.e., we have that $\bm{\theta}_\texttt{MAP}$ maximizes the conditional posterior $p(\bm{\theta}|\obs,\sigma)$ for any $\sigma$. See App. \ref{AppOptimi} for further details.
\newline
Furthermore, the variance of the conditional posterior $p(\bm{\theta}|\obs, \sigma)$ grows when  $\sigma$ increases. In this sense, with larger $\sigma$, the density $p(\bm{\theta}|\obs, \sigma)$ is `wider', hence it is easier for Monte Carlo methods to explore the space (namely, we have a {\it tempering effect}). 
Based on these considerations, we can run Monte Carlo schemes (specifically IS algorithms) on $p(\bm{\theta}|\obs, \sigma_0)$ with a large value $\sigma_0$ for estimating $\bm{\theta}_\texttt{MAP}$ more efficiently.
Furthermore, apart from estimating $\bm{\theta}_\texttt{MAP}$, we are also interested in studying the conditional posterior  $p({\bm \theta}|{\bf y},\sigma_{\texttt{ML}})$ where 
$$
{\sigma}_{\texttt{ML}}=\arg\max_{\sigma}  \ell({\bf y}|{\bm \theta}_{\texttt{MAP}},\sigma).
$$
The value ${\sigma}_{\texttt{ML}}$ can be obtained in closed-form (see App. \ref{AppOptimi}). In fact, for any $\bm{\theta}$, we have
\begin{eqnarray}
	\ell({\bf y}| {\bm \theta}, \sigma)
	&\propto& \left(\frac{1}{\sigma^2}\right)^{\frac{K}{2}} \exp\left(-\frac{||{\bf y}- {\bf f}({\bm \theta})||^2}{2\sigma^2}\right),
\end{eqnarray}
has the form of an {\it Inverse Gamma} density for $\sigma^2$, and it has  a {\it unique} mode at
$\sqrt{\frac{1}{K}||{\bf y}- {\bf f}({\bm \theta})||^2}$, {considering $\bm{\theta}$ is a fixed value}.
Hence,  finally we have 
$$
{\sigma}_{\texttt{ML}}=\sqrt{\frac{1}{K}||{\bf y}- {\bf f}({{\bm \theta}}_{\texttt{MAP}})||^2}.
$$
This can serve as a point estimator of the noise power in the system, and also as possible value to stop the tempering of the conditional posterior, as we show in the following section.

\subsection{Suggested iterative approach} 

Consider we start with a large value $\sigma_0$, which can be viewed as a coarse approximation of $\sigma_{\texttt{ML}}$, so we denote it $\sigma_0 = \widehat{\sigma}_\texttt{ML}^{(0)}$. Let $\widehat{\bm{\theta}}_\texttt{MAP}^{(1)}$ denote {an estimator} of $\bm{\theta}_\texttt{MAP}$ obtained by working w.r.t. $p(\bm{\theta}|\obs,\widehat{\sigma}_\texttt{ML}^{(0)})$. We use this current estimation to obtain the next value of $\sigma$, i.e.,  $\widehat{\sigma}_\texttt{ML}^{(1)} = \sqrt{\frac{1}{K}||\obs-{\bf f}(\widehat{\bm{\theta}}_\texttt{MAP}^{(1)})||^2}$. In general, $\widehat{\sigma}_\texttt{ML}^{(1)}$ will be a better estimator of $\sigma_\texttt{ML}$ than $\widehat{\sigma}_\texttt{ML}^{(0)}$.
We can iterate this procedure: for $t=1,\dots,T$:
\begin{itemize}
	\item[1] Estimate $\widehat{\bm{\theta}}_\texttt{MAP}^{(t)}$ by Monte Carlo (e.g., by an IS scheme) working with respect to $p(\bm{\theta}|\obs,\widehat{\sigma}_\texttt{ML}^{(t-1)})$.
	\item[2] Compute 
	$$
	\widehat{\sigma}_\texttt{ML}^{(t)} = \sqrt{\frac{1}{K}||\obs-{\bf f}(\widehat{\bm{\theta}}_\texttt{MAP}^{(t)})||^2}.
	$$
\end{itemize}
With this iterative scheme, we have that $\widehat{\sigma}_\texttt{ML}^{(T)}\rightarrow \sigma_\texttt{ML}$ as $T$ grows, hence we eventually perform IS with respect to the density of interest $p(\bm{\theta}|\obs,\sigma_\texttt{ML})$. Furthermore, a non-increasing sequence of values $\widehat{\sigma}_\texttt{ML}^{(0)} \geq \widehat{\sigma}_\texttt{ML}^{(1)} \geq \dots \geq \widehat{\sigma}_\texttt{ML}^{(T)}$ is produced, which facilitates the estimation of $\bm{\theta}_\texttt{MAP}$, and ensures the IS estimation of $p(\bm{\theta}|\obs,\sigma_\texttt{ML})$ is performed efficiently by using the set of intermediate, tempered (i.e., wider) distributions $p(\bm{\theta}|\obs,\widehat{\sigma}_\texttt{ML}^{(t)})$ for $t=0,1,...,T$. Finally, a particle approximation of $p(\bm{\theta}|\obs,\widehat{\sigma}_\texttt{ML}^{(T)})$ is obtained, i.e., 
$$
p(\bm{\theta}|\obs,\widehat{\sigma}_\texttt{ML}^{(T)})=\sum_{t=1}^T \sum_{n=1}^N \widetilde{w}_{t}^{(n)} \delta({\bm \theta}-{\bm \theta}_t^{(n)}),
$$
where $\sum_{t=1}^T\sum_{n=1}^N \widetilde{w}_{t}^{(n)}=1$.{\footnote{{Note that $\widetilde{w}^{(n)}_t$ are the final corrected weights obtained at end of the algorithm (see Table \ref{AIS_AutoTemp}).}}} 
\section{Automatic Tempering Adaptive Importance Sampling  (ATAIS)}\label{ATAIS_sect}
In this section, we describe an adaptive importance sampler with an {\it automatic tempering} approach which follows the procedure given above.
At each iteration $t$ of the algorithm, we have an  ML approximation of $\sigma$, i.e., $\widehat{\sigma}_{\texttt{ML}}^{(t-1)}$. 
 Considering Eq. \eqref{EstaEslaConDMargPost}, we define the unnormalized {\it tempered conditional posterior} at the $t$-th iteration, 
\begin{eqnarray}
\pi_t({\bm \theta})=
 \ell(\obs|\boldsymbol{\theta},\widehat{\sigma}_{\texttt{ML}}^{(t-1)}) g_\theta\mathbf{(\boldsymbol{\theta})},
\label{eq:bayes_3}
\end{eqnarray}
where we assume $g_\theta(\bm{\theta})\propto 1$ in ${\bm \Theta}$. For other generic choice of $g_\theta(\bm{\theta})$, see the discussion in Section \ref{GenPriorG}.
At each iteration, we consider $p(\bm{\theta}|\obs,\widehat{\sigma}_\texttt{ML}^{(t-1)}) \propto \pi_t(\bm{\theta})$ as the target distribution.
The dependence on the iteration $t$ is due to $\widehat{\sigma}_{\texttt{ML}}^{(t)}$ varies with $t$. The ATAIS algorithm is outlined in Table \ref{AIS_AutoTemp}. 
The resulting scheme is an adaptive IS algorithm which combines sampling schemes and stochastic optimization. 
It is important to remark that, if $\widehat{\sigma}_{\texttt{ML}}^{(0)}$ is bigger than the true ML value, we generate a non-increasing sequence of $\widehat{\sigma}_{\texttt{ML}}^{(t)}$, i.e.,
$\widehat{\sigma}_{\texttt{ML}}^{(0)}\geq\widehat{\sigma}_{\texttt{ML}}^{(1)}\geq ... \widehat{\sigma}_{\texttt{ML}}^{(t)}\geq \widehat{\sigma}_{\texttt{ML}}^{(t+1)}$, etc. Note that this is true since we have assumed a uniform prior  $g_\theta(\bm{\theta})$. {{\footnote{{To see this, recall $\widehat{\sigma}_{\texttt{ML}} = \sqrt{\frac{1}{K}||\obs - {\bf f}(\widehat{\bm{\theta}}_\texttt{MAP})||^2}$. Improving $\widehat{\bm{\theta}}_\texttt{MAP}$ means that squared error $||\obs - {\bf f}(\widehat{\bm{\theta}}_\texttt{MAP})||^2$ is lower, which implies that $\widehat{\sigma}_{\texttt{ML}}$ always decreases (provided that we start with $\widehat{\sigma}_\texttt{ML}>\sigma_\texttt{ML}$).}}}}
\newline
\newline
{\bf IS steps.}
 A set of $N$ samples $\{{\bm \theta}_{t}^{(n)}\}_{n=1}^N$ are drawn from a (normalized) proposal density $q({\bm \theta}|{\bm \mu}_t,{\bm \Sigma}_t)$ with mean ${\bm \mu}_t$ and a covariance matrix ${\bm \Sigma}_t$. An importance weight 
 $$
 w_{t}^{(n)}=\frac{\pi_t({\bm \theta}_t^{(n)})}{q({\bm \theta}_t^{(n)}|{\bm \mu}_t,{\bm \Sigma}_t)},
 $$ 
 is assigned to each sample.
\newline
\newline
{\bf Proposal adaptation.}  A particle estimation of the conditional MAP estimator of ${\bm \theta}$ is given by
$\widehat{{\bm \theta}}_t =\arg\max\limits_{n} \pi_t({\bm \theta}_t^{(n)})$.  The value of current MAP approximation $\pi_t(\widehat{{\bm \theta}}_t)$ is then compared with the value of global MAP estimator obtained so far denoted as $\pi_{\texttt{MAP}}$. If  $\pi_t(\widehat{{\bm \theta}}_t) \geq\pi_{\texttt{MAP}}$, all the global MAP estimators are updated and the proposal pdf is moved at $\widehat{{\bm \theta}}_t$, i.e., we set
\begin{eqnarray}
\widehat{{\bm \theta}}^{(t)}_{\texttt{MAP}} =\widehat{{\bm \theta}}_t,  \quad    \pi_{\texttt{MAP}}=\pi_t(\widehat{{\bm \theta}}_t), \quad {\bm \mu}_t=\widehat{{\bm \theta}}_t.
\end{eqnarray}
Otherwise, we keep the previous values $\widehat{{\bm \theta}}_{\texttt{MAP}}^{(t)}=\widehat{{\bm \theta}}_{\texttt{MAP}}^{(t-1)} $, $\pi_{\texttt{MAP}}$ and ${\bm \mu}_t={\bm \mu}_{t-1}$.  
The covariance matrix ${\bm \Sigma}_t$ is adapted by considering the empirical covariance of the weighted samples.  Note that, we set ${\bm \mu}_t=\widehat{{\bm \theta}}_{\texttt{MAP}}^{(t)}$ instead of using the empirical mean of the samples (as in other classical AIS schemes). This is because we have noticed that this choice provides better and more robust results, specially as the dimension of the problem grows.  
\newline
\newline
{\bf Automatic tempering.} As we showed in the previous section, the current ML estimator of $\sigma$ can be obtained analytically as
\begin{equation}
\widehat{\sigma}_t=\sqrt{\frac{1}{K}||\mathbf{y} - \widehat{\bf{r}}_t||^2},
\end{equation}
where $\widehat{\bf{r}}_t={\bf f}(\widehat{{\bm \theta}}_t)$.  If the  current ML estimator $\widehat{\sigma}_t$ is smaller than the current global one $\widehat{\sigma}_{\texttt{ML}}^{(t-1)}$, i.e.,  $\widehat{\sigma}_t < \widehat{\sigma}_{\texttt{ML}}^{(t-1)}$, then we update  $\widehat{\sigma}_{\texttt{ML}}^{(t)} =\widehat{\sigma}_t$,
Otherwise, we keep the value of $\widehat{\sigma}_{\texttt{ML}}^{(t)}=\widehat{\sigma}_{\texttt{ML}}^{(t-1)}$. 
{Actually, with a uniform prior  $g_\theta(\bm{\theta})$, every time that we update $\widehat{{\bm \theta}}_\text{MAP}^{(t)}$ we also update $\widehat{\sigma}_{\texttt{ML}}^{(t)}$ (see footnote in the previous page).}
 \newline
 \newline
{\bf ATAIS outputs.} 
After $T$ iterations, a final correction of the weights is needed, i.e.,
\begin{equation}\label{aquiW}
\widetilde{w}_{t}^{(n)}=w_{t}^{(n)} \frac{\pi_{T+1}({\bm \theta}_t^{(n)})}{\pi_t({\bm \theta}_t^{(n)})},
\end{equation}
in order to obtain a particle approximation of the measure of the final conditional posterior $p(\bm{\theta}|\obs,\widehat{\sigma}_\texttt{ML}^{(T)}) \propto \pi_{T+1}(\bm{\theta})$. Thus, the algorithm  returns the final estimators $\widehat{{\bm \theta}}_{\texttt{MAP}}^{(T)}$, $\widehat{\sigma}_{\texttt{ML}}^{(T)}$, and all the weighted samples $\{{\bm \theta}_{t}^{(n)},\widetilde{w}_{t}^{(n)}\}$, for all $n=1,...,N$ and $t=1,...,T$. Other outputs can be obtained with a post-processing of the weighted samples, as shown below. Note that Eq. \eqref{aquiW} does not require any additional evaluation of the model, and the error $e_t^{(n)}=||{\bf y}- {\bf f}({\bm \theta}_t^{(n)}) ||^2$.
Moreover, we can also use $e_t^{(n)}$  and $\{{\bm \theta}_{t}^{(n)}\}$ for building a particle approximation of  any other conditional posterior $p(\bm{\theta}|\obs,\sigma)$. This allows the study of the marginal posterior $p(\sigma|\obs)$ and provides the complete Bayesian inference, as we  show in the next section.  


\begin{table}[!h]
\caption{ATAIS: AIS with automatic tempering \label{AIS_AutoTemp}}
\begin{tabular}{|p{0.95\columnwidth}|}
    \hline
\begin{enumerate}
\item {\bf Initializations:} Choose $N$, ${\bm \mu}_1$, ${\bm \Sigma}_1$,  and 
obtain an initialization for  $\widehat{\sigma}_{\texttt{ML}}^{(0)}$, and set $\pi_{\texttt{MAP}}=0$.
 \item {\bf For $t=1,\ldots,T$:}
\begin{enumerate}
\item {\bf Sampling:}
\begin{enumerate}
 \item  Draw ${\bm \theta}_{t}^{(1)},...,{\bm \theta}_{t}^{(N)} \sim q({\bm \theta}|{\bm \mu}_t,{\bm \Sigma}_t)$.
\item  Assign to each sample the weights 
  \begin{eqnarray}
w_{t}^{(n)}=\frac{\pi_t({\bm \theta}_t^{(n)})}{q({\bm \theta}_t^{(n)}|{\bm \mu}_t,{\bm \Sigma}_t)}
, \qquad n=1,...,N.
\end{eqnarray}
\end{enumerate}
\item {\bf Current maximum estimations:} 
\begin{enumerate}
\item Obtain $\widehat{{\bm \theta}}_t =\arg\max\limits_{n} \pi_t({\bm \theta}_t^{(n)})$,  and compute
$\widehat{\bf{r}}_t={\bf f}(\widehat{{\bm \theta}}_t)$ 
\item Compute $\widehat{\sigma}_t=\sqrt{\frac{1}{K}||\mathbf{y} - \widehat{\bf{r}}_t||^2}$.
\end{enumerate}
\item {\bf Global maximum estimations:}
\begin{enumerate}
\item   If  $\widehat{\sigma}_t \leq \widehat{\sigma}_{\texttt{ML}}^{(t-1)}$, then set $ \widehat{\sigma}_{\texttt{ML}}^{(t)} =\widehat{\sigma}_t$. Otherwise, set $\widehat{\sigma}_{\texttt{ML}}^{(t)}=\widehat{\sigma}_{\texttt{ML}}^{(t-1)}$. 
\item   If  $\pi_t(\widehat{{\bm \theta}}_t) \geq\pi_{\texttt{MAP}}$, then set $\widehat{{\bm \theta}}_{\texttt{MAP}}^{(t)}=\widehat{{\bm \theta}}_t$ and   $\pi_{\texttt{MAP}}=\pi_t(\widehat{{\bm \theta}}_t)$. Otherwise, $\widehat{{\bm \theta}}_{\texttt{MAP}}^{(t)}=\widehat{{\bm \theta}}_{\texttt{MAP}}^{(t-1)}$ and keep the value of $\pi_\texttt{MAP}$.
\end{enumerate}
\item {\bf Adaptation:} Set 
\begin{eqnarray}
{\bm \mu}_t&=&\widehat{{\bm \theta}}^{(t)}_{\texttt{MAP}}, \\
{\bm \Sigma}_t&=&\sum_{n=1}^N \bar{w}_t^{(n)} ({\bm \theta}_t^{(n)}-{\bar {\bm \theta}}_t)^{\top} ({\bm \theta}_t^{(n)}-{\bar {\bm \theta}}_t) + {\epsilon}{\bf I}_M,
\end{eqnarray}
where  $\bar{w}_t^{(n)} \frac{w_t^{(n)}}{\sum_{i=1}^N w_t^{(i)}}$ are the normalized weights, ${\bar {\bm \theta}}_t=\sum_{n=1}^N \bar{w}_t^{(n)} {\bm \theta}_t^{(n)}$ and {$\epsilon>0$ is a small scalar value} .
	\end{enumerate}
\item {\bf Output:} Return the  final estimators $\widehat{{\bm \theta}}^{(T)}_{\texttt{MAP}}$, $\widehat{\sigma}^{(T)}_{\texttt{ML}}$, and all the weighted samples $\{{\bm \theta}_{t}^{(n)},\widetilde{w}_{t}^{(n)}\}$, for all $t$ and $n$, with the corrected weights
\begin{equation}
\widetilde{w}_{t}^{(n)}=w_{t}^{(n)} \frac{\pi_{T+1}({\bm \theta}_t^{(n)})}{\pi_t({\bm \theta}_t^{(n)})}.
\end{equation}
\end{enumerate} \\
\hline 
\end{tabular}
\end{table}

\subsection{For a generic prior $g_\theta(\bm{\theta})$} \label{GenPriorG}

The ATAIS algorithm is based on the fact that $\bm{\theta}_\texttt{MAP}$ does not depend on $\sigma$. This allows us to progressively estimate it by targeting the sequence of tempered posteriors $\pi_t(\bm{\theta}) \propto p(\bm{\theta}|\obs,\widehat{\sigma}^{(t)}_\texttt{ML})$ that share all the same MAP. However, in the case $g_\theta(\bm{\theta})$ is not uniform, we generally have one $\bm{\theta}_{\texttt{MAP}}(\sigma)$ for each $p(\bm{\theta}|\obs,\sigma)$, and we could have that the sequence of $\bm{\theta}_\texttt{MAP}(\widehat{\sigma}_{\texttt{ML}}^{(t)})$ will not approach $\bm{\theta}_\texttt{MAP}(\sigma_{\texttt{ML}})$. 
\newline
If the data are informative and the prior $g_\theta(\bm{\theta})$ is chosen such it is vague  with respect to the likelihood, the position of $\bm{\theta}_{\texttt{MAP}}(\sigma)$ is not very sensitive to the value of $\sigma$. Namely, we have $\bm{\theta}_\texttt{MAP}(\widehat{\sigma}_{\texttt{ML}}^{(1)}) \approx \bm{\theta}_\texttt{MAP}(\widehat{\sigma}_{\texttt{ML}}^{(2)}) \approx \dots \approx \bm{\theta}_\texttt{MAP}({\sigma}_{\texttt{ML}})$, and hence our algorithm can be applied in this context. 
{ When the data are not informative, we should use an even more vague  prior (i.e. wider than the likelihood function) in order to maintain the usefulness of the algorithm.}


\newpage
\section{Complete Bayesian inference with ATAIS}\label{SectSuperFer}

Let us assume we have a proper prior $g_\theta(\bm{\theta})$ and we introduce another proper prior $g_\sigma(\sigma)$ for $\sigma$.
The outputs of the ATAIS algorithm can serve to approximate the normalizing constant of the joint posterior $p(\bm{\theta},\sigma|\obs) \propto \ell(\obs|\bm{\theta},\sigma)g_\theta(\bm{\theta})g_\sigma(\sigma)$, i.e., the so-called marginal likelihood or Bayesian model evidence, given by
\begin{equation}\label{aquiZconZ}
Z=\int_{\mathbb{R}^+}\int_{\bm{\Theta}} \ell(\obs|\bm{\theta},\sigma)g_\theta(\bm{\theta})g_\sigma(\sigma) d\bm{\theta}d\sigma
=\int_{\mathbb{R}^+}Z(\sigma)g_\sigma(\sigma)d\sigma,
\end{equation}
where we have denoted $Z(\sigma) = \int_{\bm{\Theta}} \ell(\obs|\bm{\theta},\sigma)g_\theta(\bm{\theta} d\bm{\theta}$, usually called  conditional marginal likelihood. The quantity $Z$ is useful for model selection purposes.  Furthermore, a complete Bayesian study of the joint posterior $p({\bm \theta}, \sigma|\obs)$ can be provided as well.
\newline
\newline
{\bf Approximation of $Z(\sigma)=p({\bf y}|\sigma)$.} 
After the $T$ iterations of ATAIS, we can also approximate  the conditional marginal likelihood  $Z(\sigma)=p({\bf y}|\sigma)$ without additional evaluations of the target function. Indeed, saving the error values at each particle obtained for the computation of the likelihood function during ATAIS, 
$$
e_t^{(n)}=||{\bf y}- {\bf f}({\bm \theta}_t^{(n)})||^2,
$$
we can compute the IS weights,
\begin{equation}\label{Rho_aqui_primero}
\rho_{t}^{(n)}(\sigma)=\frac{\frac{1}{(2\pi\sigma^2)^{\frac{K}{2}}} \exp\left(-\frac{e_t^{(n)}}{2\sigma^2}\right)  g_\theta({\bm \theta}^{(n)}_t)}{q({\bm \theta}_t^{(n)}|{\bm \mu}_t,{\bm \Sigma}_t)},
\end{equation} 
for a generic value of $\sigma$
Thus, the IS estimator of the  conditional marginal likelihood  $Z(\sigma)$ is given by the arithmetic mean of the weights $ \rho_{t}^{(n)}(\sigma)$,
\begin{equation}
	\widehat{Z}(\sigma)=\widehat{p}({\bf y}|\sigma)=\frac{1}{NT} \sum_{t=1}^T\sum_{n=1}^N  \rho_{t}^{(n)}(\sigma).
\end{equation}
{\bf Approximation of $Z$.}
Drawing $\sigma^{(r)} \sim g_\sigma(\sigma)$, for $r=1,...,R$, (or considering a deterministic grid, {e.g., as a Riemannian integration}), we can approximate the global marginal likelihood $Z$ by applying simple Monte Carlo to the integral in Eq. \eqref{aquiZconZ},
\begin{equation}
\widehat{Z}=\frac{1}{R}\sum_{r=1}^R \widehat{Z}(\sigma^{(r)}).
\end{equation}
{\bf Approximation of $p(\sigma|\obs)$.} An approximation of the marginal posterior $p(\sigma|\obs) = \frac{p(\obs|\sigma)g_\sigma(\sigma)}{p(y)}$ can be also obtained as
\begin{align}\label{ApproxMargPostEq}
p(\sigma|\obs) \approx
\widehat{p}(\sigma|\obs) = \frac{\widehat{Z}(\sigma)g_\sigma(\sigma)}{\widehat{Z}},
\end{align}
which can be used to approximate, e.g., the MAP of $p(\sigma|\obs)$ by 
$
\sigma_{\texttt{MAP-marg}} \approx \arg\max_\sigma\widehat{Z}(\sigma)g_\sigma(\sigma).
$
Other different moments of $p(\sigma|\obs)$ can be computed by a deterministic quadrature (since the problem is now one-dimensional) or applying noisy Monte Carlo approaches.
\newline
{\bf Complete Bayesian analysis.}   We can approximate the integral of interest as
\begin{align}\label{INT_gen2}
 I&= \int_{\mathbb{R}^+} \int_{{\bm \Theta}} h({\bm \theta},\sigma)  p({\bm \theta},\sigma|{\bf y}) d{\bm \theta} d\sigma, \\
 &=  \int_{\mathbb{R}^+} \int_{{\bm \Theta}} h({\bm \theta},\sigma)  p({\bm \theta}|{\bf y},\sigma)p(\sigma|{\bf y}) d{\bm \theta} d\sigma \\
& \approx   \frac{1}{J}\sum_{j=1}^J \sum_{t=1}^T \sum_{n=1}^N \bar{\rho}_{t}^{(n)}(\sigma^{(j)}) h({\bm \theta}_{t}^{(n)},\sigma^{(j)}),
\end{align}
where 
\begin{align}\label{full_Bayes_weights}
	\bar{\rho}_{t}^{(n)}(\sigma^{(j)})=\frac{\rho_{t}^{(n)}(\sigma^{(j)})}{ \sum_{\tau=1}^T \sum_{i=1}^N \rho_{\tau}^{(i)}(\sigma^{(j)})},
\end{align}
and $\sigma^{(j)}$ are generated by applying a noisy MCMC with invariant density $\widehat{p}(\sigma|\obs)\propto \widehat{Z}(\sigma)g_\sigma(\sigma)$. Note that the samples ${\bm \theta}_{t}^{(n)}$ do not depend on the index $j$ (they do not change) since we are {\it recycling} the particles generated by ATAIS and reusing evaluations $e_t^{(n)}=||{\bf y}- {\bf f}({\bm \theta}_t^{(n)})||^2$.

\section{Simulations}
\label{sec:Simul}
We test the proposed scheme in two numerical examples. The first numerical experiment is a simple bidimensional example  (which is easy to be reproduced). The second experiment considers a  real-world application, i.e., a radial velocity models of exoplanet systems which is often employed in astronomy applications (with a dimension of the inference problem of $6$ and $11$). 

\subsection{First numerical analysis}
\begin{figure*}[!h]
\begin{center}
\centerline{
\subfigure[$\sigma=20$.]{\includegraphics[width=0.3\columnwidth]{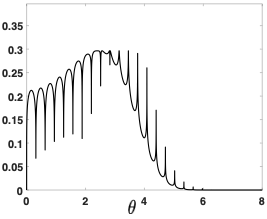}}
\subfigure[$\sigma=10$.]{\includegraphics[width=0.3\columnwidth]{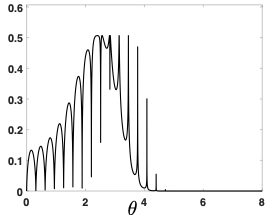}}
\subfigure[$\sigma=\sigma_{\texttt{true}}=4$.]{\includegraphics[width=0.3\columnwidth]{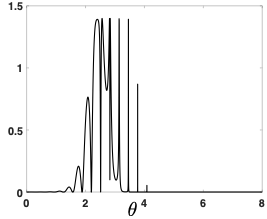}}
}
\vspace{-0.3cm}
\caption{Conditional posteriors corresponding to different values of $\sigma$: {\bf (a)} $\sigma=20$, {\bf (b)} $\sigma=10$, \newline {\bf (c)} $\sigma=\sigma_{\texttt{true}}=4$. 
}
\label{figEX1SeqTarget}
\end{center}
\end{figure*}

\begin{figure*}[!h]
\begin{center}
\centerline{
\subfigure[Joint posterior.]{\includegraphics[width=0.3\columnwidth]{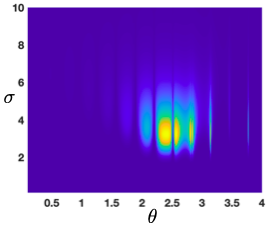}}
\subfigure[Marginal posterior $p(\theta|\obs)$.]{\includegraphics[width=0.3\columnwidth]{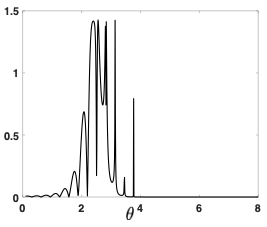}}
\subfigure[Marginal posterior $p(\sigma|\obs)$.]{\includegraphics[width=0.3\columnwidth]{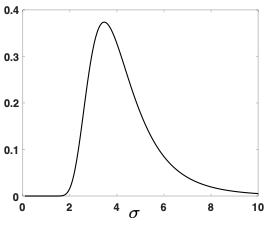}}
}
\vspace{-0.3cm}
\caption{The bidimensional joint posterior $p(\theta,\sigma|\obs)$ and the two marginal posteriors $p(\theta|\obs)$, $p(\sigma|\obs)$ in Eq. \eqref{EqMargPost}, computed by using a thin grid approximation.}
\label{figEX1SeqTarget2}
\end{center}
\end{figure*}

\begin{figure*}[!h]
	\begin{center}
		\centerline{
			\subfigure[Histogram with $2\cdot 10^6$ samples.]{\includegraphics[width=0.3\columnwidth]{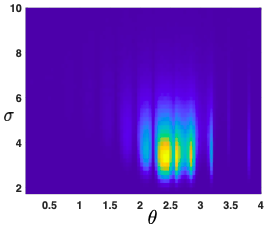}}		
			\subfigure[$10^4$ samples by ATAIS.]{\includegraphics[width=0.298\columnwidth]{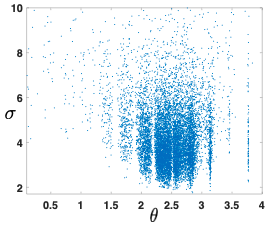}}	
			\subfigure[Histogram with $2\cdot 10^6$ samples.]{\includegraphics[width=0.3\columnwidth]{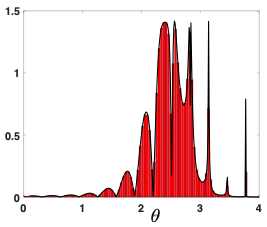}}
		}
	\vspace{-0.3cm}
		\caption{{Approximations obtained with ATAIS. {\bf (a)-(b)} Joint posterior $p(\bm{\theta},\sigma|\obs)$: {\bf (a)} by an histogram with $2 \cdot 10^6$ samples;  {\bf (b)} $10^4$ samples from joint posterior obtained by ATAIS. {\bf (c)} Approximation by an histogram with  $2\cdot 10^6$ samples, of the marginal posterior $p(\bm{\theta}|\obs)$ .}}
		\label{fig_lucatitular}
	\end{center}
\end{figure*}

\begin{figure*}[!h]
\begin{center}
\centerline{
\subfigure[ $\widehat{\sigma}_{\texttt{ML}}^{(t)}$ vs $t$.]{\includegraphics[width=0.3\columnwidth]{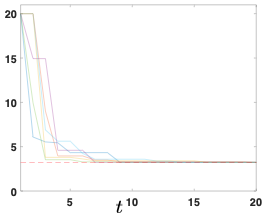}}
\hspace{0.6cm}
\subfigure[ $p(\sigma|\obs)$ for different $N$ ($T=10$).]{\includegraphics[width=0.3\columnwidth]{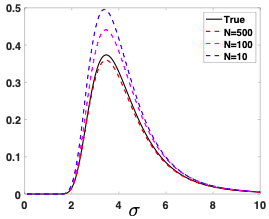}}
}
\caption{{\bf (a)} The ML estimation $\widehat{\sigma}_{\texttt{ML}}^{(t)}$ (different runs) versus the number of iterations $t$, with $N=5$.  {\bf (b)} The true marginal posterior $p(\sigma|\obs)$ and different approximations, in one specific run, $\widehat{p}(\sigma|\obs)$ obtained as in Eq. \eqref{ApproxMargPostEq} with different $N\in\{10,100,500\}$ and $T=10$ (hence, the total number of samples are $NT$).}
\label{figEX1SeqTarget3}
\end{center}
\end{figure*}

For the sake of simplicity, let us consider $\theta\in\mathbb{R}$ and an observation model given by the equation
$$
y_k= \theta^2+\log(|\sin(10\theta)|)+ v_k,
$$
so that $f(\theta)=\theta^2+\log(|\sin(10\theta)|)$, and $v_k \sim \mathcal{N}(0,\sigma^2)$. We consider  $\theta_{\texttt{true}}=2.5$, and $\sigma_{\texttt{true}}=4$.  We generate $K=8$ observations from the model above.  We  also consider a uniform prior for $\theta$ in ${(}0,20]$. The conditional posterior $p(\theta|\obs,\sigma_{\texttt{true}})$ is shown in Figure \ref{figEX1SeqTarget}(c). We can observe that $p(\theta|\obs,\sigma_{\texttt{true}})$  is highly multimodal. Figure \ref{figEX1SeqTarget} also depicts the conditional posteriors $p(\theta|\obs,\sigma)$ with $\sigma\in \{10,20\}$. Considering also  a uniform prior over $\sigma$ in ${(}0,20]$, we have also a bidimensional joint posterior over $[\theta,\sigma]$, which is depicted in Fig. \ref{figEX1SeqTarget2}(a).
\newline
In this bidimensional example, it is possible to obtain the ground-truths using an expensive thin grid.  
{We show the ground-truths of the different pdfs in Table \ref{tab:ground_truths}.}
{ Moreover, the true value of the complete evidence $Z=p({\bf y})=1.5983 \cdot 10^{-9}$.}
Since the prior over $\sigma$ is uniform, the maximum likelihood of $\sigma$ is $\sigma_{\texttt{ML}}=\sigma_{\texttt{MAP-joint}}=3.23$.
The two marginal posteriors are shown in Figures \ref{figEX1SeqTarget2}(b)-(c).
\begin{table}[!h]
{
	\caption{{\footnotesize Summary of pdfs and ground-truths for the first numerical experiment.} }   
\label{tab:ground_truths} 
\vspace{-0.3cm}
\begin{center}
\begin{tabular}{|c|c|c|c|} 
	\hline
	Pdf  & Expectation & Variance  & MAP \\
	\hline
	\hline
	$p(\bm{\theta}|\obs,\sigma_\texttt{ML})$ & 2.48 & 0.11 & 2.56 \\
	$p(\sigma|\obs)$ & 4.32 & 2.43 & 3.46 \\
	$p(\bm{\theta}|\obs)$ & 2.46 & 0.18 & 2.56 \\
	\hline
\end{tabular}
\end{center}
}
\end{table}
\newline
 We apply ATAIS with the goal of estimating the expected value and the variance of the posterior density with respect to $\theta$. We consider a Gaussian proposal $q(\theta|\mu_t,\lambda_t)$ with $\mu_0=10$ and a starting variance of $\lambda_0=4$.  Note that $\mu_0$ is located in a region that does not contain modes.  We also start with $\widehat{\sigma}_{\texttt{ML}}^{(0)}=20$ and $\pi_{\texttt{MAP}}=0$ (initial conditions). 
 The Mean Square Error (MSE) of ATAIS, averaged over $500$ runs, in estimation of different moments and modes as function of $N$ (and with $T=10$), is given in Table \ref{tab:MSE_ATIS}. The ML estimation $\widehat{\sigma}_{\texttt{ML}}^{(t)}$ as function of the iteration $t$ (with $N=5$) for different runs, is given in Figure \ref{figEX1SeqTarget3}(a). The approximation of the marginal posterior ${p}(\sigma|\obs)$, denoted $\widehat{p}(\sigma|\obs)$, is  obtained as in Eq. \eqref{ApproxMargPostEq} in one specific run, with different $N\in\{10,100,500\}$ and $T=10$.
 {
The approximations of the joint posterior $p(\bm{\theta},\sigma|\obs)$ and the marginal posterior $p(\bm{\theta}|\obs)$, obtained by resampling the particles according to the normalized weights in Eq. \eqref{full_Bayes_weights} and \eqref{Rho_aqui_primero}, are shown in Figure \ref{fig_lucatitular}, i.e., using a sampling importance resampling procedure. For more details, see \cite{Rubin88} and \cite[Chapter 24]{gelman2004applied}.
}
 
  
\begin{table}[!h]
   \caption{{\footnotesize MSE of ATAIS (averaged over $500$ runs), in the estimation of the evidence, different moments and modes as function of $N$ and $T=10$.} }   
   \label{tab:MSE_ATIS} 
   \centering
   \vspace{0.2cm}
   \begin{tabular}{|c|c|c|c|c|c|} 
      \hline
     {\bf Value} & $N=10$ & $N=100$  &  $N=1000$ &  $N=5000$ & {{\bf Ground-truths}}\ \\
      \hline
      \hline
      $E[\theta|\obs,\sigma_{\texttt{ML}}]$                &  0.0311          & 0.0098 & 0.0034  & 0.0024  & {2.48}\\
      $\mbox{var}[\theta|\obs,\sigma_{\texttt{ML}}]$       &     0.0474   &  0.0370   & 0.0298   & 0.0201  & {0.11}\\
       $\theta_{\texttt{MAP}}$  & 0.0410   & 0.0337 & 0.0285 &  0.0127 & {2.56} \\
      $E[\sigma|\obs]$ &  0.9233   &  0.0785  &  0.0097 & 0.0023 & {4.32} \\
      $\mbox{var}[\sigma|\obs]$ & 6.1869 & 0.2640 & 0.0035 &  0.0010 & {2.43}\\
       $\sigma_{\texttt{MAP-marg}}$ & 0.0056 & 0.0004& 0.0001 &  $3\cdot 10^{-5}$ & {3.46}\\ 
         $\sigma_{\texttt{ML}}$ & $8\cdot 10^{-5}$ & $2\cdot 10^{-5}$ & $5\cdot 10^{-7}$ & $6\cdot 10^{-9}$ & {3.23}\\ 
        { $Z=p({\bf y})$ } &   {$2 \cdot 10^{-18}$}  &  {$1.8 \cdot 10^{-20}$}  &  {$1.4 \cdot 10^{-20}$}   &  {$3.6 \cdot 10^{-22}$}  & {$1.6 \cdot 10^{-9}$}\\
      \hline
   \end{tabular}
\end{table}

\newpage

\subsection{Radial velocity curves of exoplanets and binary systems}
\label{sec:Exo}

In this example, we consider an application in an astronomical model.
In recent years, the problem of revealing objects orbiting other stars has acquired large attention. Different techniques have been proposed to discover exo-objects but, nowadays, the radial velocity technique is still the most used \cite{Gregory2011,Barros2016,Affer2019,Trifonov2019}. The problem consists in fitting a model (the so-called radial velocity curve) to data acquired at different moments spanning during long time periods (up to years). The model is highly non-linear and it is costly in terms of computation time (specially, for certain sets of parameters). Obtaining a value to compare to a single observation involves numerically integrating a differential equation in time or an iterative procedure for solving to a non-linear equation. Typically, the iteration is performed until a threshold is reached or $10^6$ iterations are performed. The problem of radial velocity curve fitting is applied in several related applications.

\noindent{\bf Observation model - likelihood.} When analysing radial velocity data of an exoplanetary system, it is commonly accepted that the \emph{wobbling} of the star around the centre of mass is caused by the sum of the gravitational force of each planet independently and that they do not interact with each other. Each planet follows a Keplerian orbit and the radial velocity of the host star is given by
\begin{equation}
{y}_{r,t} = V_0 + \sum\limits_{i = 1}^{S} A_i \left[ \cos \left( {u}_{i,t} + \omega_{i} \right) + e_i \cos \left( \omega_{i} \right) \right] +\xi_t,
\label{eq:rv}
\end{equation}
with $t=1,\ldots,T$ and $r=1,\ldots,R$. { In this equation, $A_i$ is the amplitude of the curve, $w_i$ is the argument of perigee and $e_i$ is the eccentricity of the orbit, of the $i$-th planet.
The parameter $V_0$ represents the mean velocity, and is common for all the planets.
}
The number of objects in the system is $S$, that is consider known in this experiment (for the sake of simplicity). Both ${y}_{r,t}$, ${u}_{i,t}$ depend on time $t$, and then $\xi_t$ is a Gaussian noise perturbation with variance $\sigma^2$.
The likelihood  function is defined by \eqref{eq:rv} and some indicator variables described below. 
 The angle ${u}_{i,t}$ is 
the true anomaly of the planet $i$ and it can be determined from
\begin{equation}
\frac{d{u}_{i,t}}{dt} = \frac{2\pi}{P_i} \frac{\left( 1 + e_i \cos {u_{i,t}} \right)^2}{\left( 1 - e_i \right)^\frac{3}{2}}
\label{eq:trueanomaly}
\end{equation}
This equation has analytical solution. As a result, the true anomaly $u_t$ can be determined from the mean anomaly $M$. However, the analytical solution contains a non linear term that needs to be determined by iterating. First, we define the mean anomaly $M_{i,t}$ as
\begin{equation}
M_{i,t} = \frac{2\pi}{P_i} \left( t - \tau_i \right),
\label{eq:meananomaly}
\end{equation}
where $\tau_i$ is the time of periastron passage of the planet $i$ and $P_i$ is the period
of its orbit. Then, through the Kepler's equation, 
\begin{equation}
M_{i,t} = E_{i,t} - e_i \sin E_{i,t},
\label{eq:kepler}
\end{equation}
where $E_{i,t}$ is the eccentric anomaly. Equation~\eqref{eq:kepler} has no analytic solution and it must be solved by an iterative procedure. A Newton-Raphson method is typically used to find the roots of this equation \cite{Press2002}. For certain sets of parameters, this iterative procedure can be particularly slow. We also have
\begin{equation}
\tan \frac{u_{i,t}}{2} = \sqrt{ \frac{1 + e_i}{1 - e_i}} \, \tan \frac{E_{i,t}}{2}, 
\label{eq:eccentricanomaly}
\end{equation}
%
The variable of interest ${\bm \theta}$ is then the vector 
\begin{equation}
{\bm \theta}= [V_0, A_{1}, \omega_{1}, e_1, P_1, \tau_1, \ldots, A_{S}, \omega_{S}, e_S, P_S, \tau_S],
\end{equation}
Then, for a single object (e.g., a planet or a natural satellite), the dimension of ${\bm \theta}$ is $M = 5+1=6$, with two objects the dimension of ${\bm \theta}$ is is $M = 11$ etc.

This example consists in a synthetic radial velocity curve of a planetary system with one planet or two planets (i.e., $S=1$ or $S=2$).  More specifically, we generate simulated data with a model with two planets. The orbital parameters of the planets are listed in Table~\ref{tab:exoplanets}, where $P$ is the period of the orbit, $A$ is the amplitude of the curve, $e$ is the eccentricity of the orbit, $\omega$ is the argument of perigee and $\tau$ is the last periastron passage. A mean velocity $V_0 = 5$~m\,s$^{-1}$ is assumed. A Gaussian noise perturbation is added with a standard deviation $\sigma = 3$~m\,s$^{-1}$. To simulate observations, a total of $K=120$ data points are selected from three, random time periods (and two planets in the system). Note that the amplitude of the radial velocity curve of the second planet is close to the noise level. We run ATAIS and a standard AIS scheme with the model with one planet and with the model with two planets. The purpose of this simulation is to check the ability of the method to detect the two planets (by approximating the model evidence). 

\begin{table}[!h]
   \caption{Main orbital parameters of the two exoplanets in the simulation.}   
   \label{tab:exoplanets} 
   \centering
   \begin{tabular}{lll} 
      \hline
      Parameter  & Planet~1 & Planet~2 \\
      \hline
      $P$                & 15 d                  & 115 d \\
      $A$                & 25 m\,s$^{-1}$ & 5 m\,s$^{-1}$ \\
      $e$                &   0.1                   & 0.0 \\
      $\omega$      & 0.61 rad         &  0.17 rad \\ 
      $\tau$            &   3 d                  &   24 d \\
      \hline
   \end{tabular}
\end{table}

\begin{figure*}[!h]
\centerline{
\includegraphics[width=0.5\columnwidth]{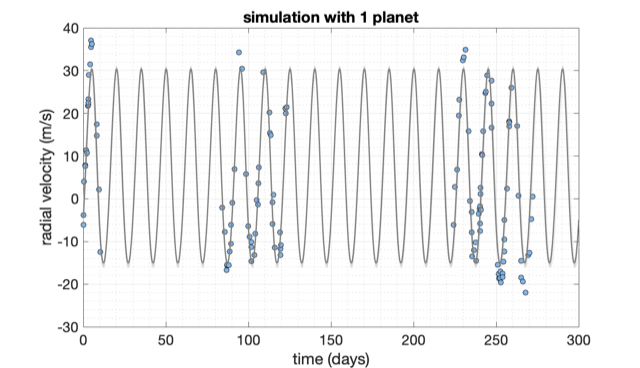}
\includegraphics[width=0.5\columnwidth]{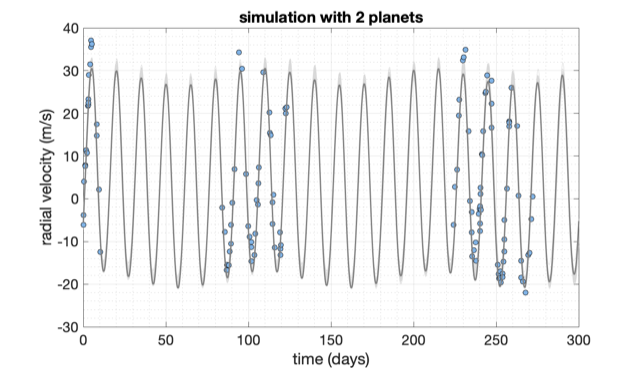}
}
\caption{Comparison of the results of the ATAIS algorithm with the simulations (blue dots).
              Left panel shows, in grey, the radial velocity curve for $\widehat{{\bm \theta}}_{\texttt{MAP}}$ 
              using a model with one planet. Right panel is like left panel but considering a model 
              with two planets. 
}
\label{fig:simulExo}
\end{figure*}

We apply ATAIS and a standard AIS scheme \cite{Bugallo15} over the space $[{\bm \theta},\sigma]$ for approximating the model evidence $Z=p({\bf y})$ (marginal likelihood) of both models (one planet or two planets) with the given data (generated considering two planets). Uniform priors are considered for each parameter: $P\in[0, 365]$, $A\in[-20,20]$,  $e\in[0,1]$, $\omega\in [0,2\pi]$,  and $\tau\in[0,50]$ (moreover, $\sigma\in[0,30]$ for the  standard AIS scheme).
The ATAIS algorithm  and the standard AIS scheme have been run with $N=10^6$  and $T=50$ iterations for both, the model with one and two planets. In both cases, we consider the same Gaussian proposal with a starting standard deviation of $5$ for each component (note that the standard AIS scheme works in higher dimensional space due the inference over $\sigma$).    
To decide which model is more probable, the model evidence $Z$ of each model  is estimated. More specifically, we  approximate the one-planet model $\widehat{Z}_1=\widehat{p}_1(\obs)$, and the two-planets model $\widehat{Z}_2=\widehat{p}_2(\obs)$ with the ATAIS algorithm  and the standard AIS scheme. When $\widehat{Z}_1 >\widehat{Z}_2$ we select the first model otherwise if, $\widehat{Z}_1 <\widehat{Z}_2$, we select the second one. The true model is the two-planets model, since the simulated data were generated from that model.
{ After $500$ independent runs, the percentage of correct detection of the true model for ATAIS is $\approx 98\%$, whereas  with the standard AIS scheme is only $\approx 56\%$. This is due to the difficulty of making inference jointly over $[\bm{\theta},\sigma]$.  Let us denote the Bayesian factor as $B=Z_2/Z_1$. In ATAIS, the expected value of the  ratio between the model evidences (averaged over the $500$ runs) is $E[B] \approx 5\cdot 10^3$ with a relative variance of $\frac{E[(B-E[B])^2]}{E[B]^2} \approx 0.04$. In the case of the standard AIS, we have $E[B] \approx 16.32$ and   $\frac{E[(B-E[B])^2]}{E[B]^2} \approx 0.15$. Therefore, for ATAIS, the model with two planets is clearly more probable than the model with one planet.} 

\begin{figure}[!h]
\centerline{
\includegraphics[width=0.4\columnwidth]{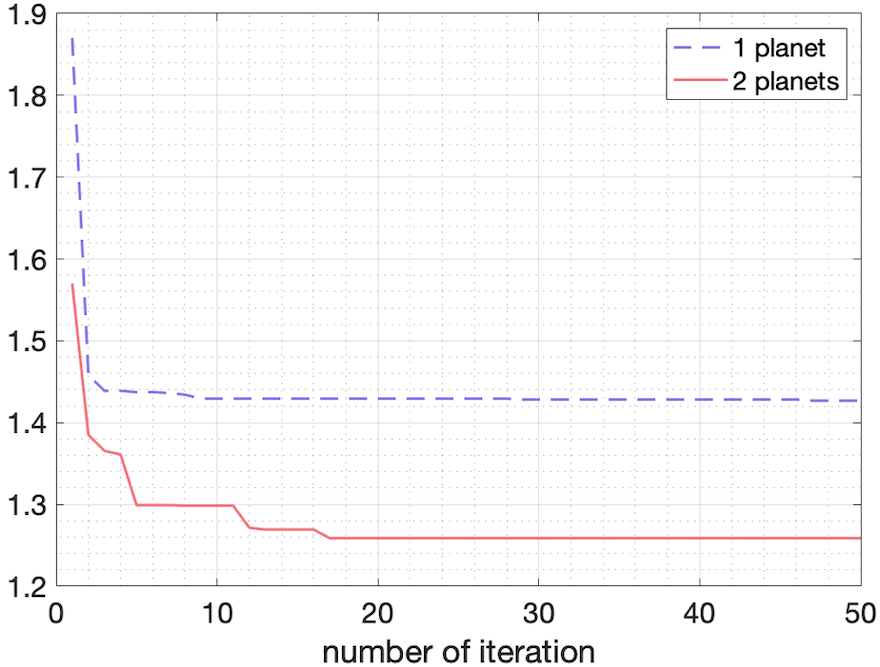}
}
\caption{Evolution of the tempering parameter $\widehat{\sigma}_{\texttt{ML}}^{(t)}$ {. We decide $\widehat{\sigma}_{\texttt{ML}}^{(0)}=50$ as starting value (the figure shows from $t=1$), which an arbitrary high value to help the exploration in the first iteration. However, after the first iteration, the algorithm is able to obtain reasonable values of $\widehat{\sigma}_{\texttt{ML}}^{(1)}$.  }
              The dashed line is the evolution for the model with one planet. The continuous line is the 
              evolution of the two-planets model. 
}
\label{fig:gamma}
\end{figure}

The fitted curves, corresponding to the vector of parameters $\widehat{{\bm \theta}}_{\texttt{MAP}}$ obtained with ATAIS, are shown in Fig.~\ref{fig:simulExo}.
 From the figure, it is not clear which model fits better the simulated observations (blue points), although the model with two planets seems to fit better the observations in the time period from  200 to 300 days. The values of $\widehat{{\bm \theta}}_{\texttt{MAP}}$, obtained in one specific run by ATAIS, is given in Table~\ref{tab:MAP}. We notice that $\omega$ and $\tau$ are highly correlated and more iterations may be needed to obtain the actual global maximum, but the remaining parameters obtained from $\widehat{{\bm \theta}}_{\texttt{MAP}}$ are similar to the simulated values. 
 In addition, the amplitude of the curve of the second planet is close to the intensity of the noise, what makes difficult to derive the best fit for that planet. Summarizing, our results show the method is able to discriminate between a model with one planet (with 6 dimensions of the inference problem) and a model with two planets (with 11 dimensions of the inference problem), for this particular simulation.  Finally, the evolution of the automatic tempering parameter $\widehat{\sigma}_{\texttt{ML}}^{(t)}$ is shown in Fig.~\ref{fig:gamma}. The dashed line is the evolution of $\widehat{\sigma}_{\texttt{ML}}^{(t)}$ for the single-planet model, whereas the continuous line is the evolution of $\widehat{\sigma}_{\texttt{ML}}^{(t)}$ for the model with two planets. In this second model, the tempering parameter reaches a smaller value,  as expected.

\begin{table}[!h]
   \caption{The value of $\widehat{{\bm \theta}}_{\texttt{MAP}}$  {and the variances of the marginal posteriors} for the 2-planets model (with $K=120$ data points). }   
   \label{tab:MAP} 
   \centering
   \vspace{0.2cm}
   \begin{tabular}{|c||c|c||c|c|} 
       \hline
   \multirow{2}{*}{ {\bf Parameter } }        &  \multicolumn{2}{c||}{{\bf  Planet~1}} &   \multicolumn{2}{c|}{{\bf  Planet~2}}  \\ 
   \cline{2-5}
 & $\widehat{{\bm \theta}}_{\texttt{MAP}}$   & {$\mbox{Var}(\theta|{\bf y})$} & $\widehat{{\bm \theta}}_{\texttt{MAP}}$    {\bf - Planet~2} &  {$\mbox{Var}(\theta|{\bf y})$}  \\
      \hline
      $P$                & 14.99 d                  &  { 0.18} & 110.39 d  &   {11.28} \\
      $K$                & 23.78 m\,s$^{-1}$  & { 0.52}  & 3.50 m\,s$^{-1}$ & { 0.44}\\
      $e$                &   0.05                      &  { 0.047} & 0.00 &  { 0.003} \\
      $\omega$      &   7.69 rad         & { 0.61} & 0.68 rad     &  { 0.82} \\ 
      $\tau$            &   6.8 d                     & { 0.76} &   7.96 d & {20.31} \\
      \hline
   \end{tabular}
\end{table}

\section{Conclusions}
We have proposed a novel AIS scheme for Bayesian inversion problems where an automatic tempering procedure is implemented (called ATAIS).
The inference of the variables of interest $\boldsymbol{\theta}$ and the noise power $\sigma^2$ is divided. A sampling strategy is considered for  $\boldsymbol{\theta}$ and an optimization approach is employed for $\sigma^2$. Thus, ATAIS  performs an iterative procedure, alternating sampling and optimization steps. Therefore, the proposed scheme deals with a sequence of tempered posteriors according to the current estimation of the noise power. We have also discussed the possibility of approximating the marginal posterior of $\sigma$ without additional evaluations of the complex model.  Furthermore, the complete Bayesian analysis regarding the complete joint posterior is possible as discussed in Section \ref{SectSuperFer}, again without any additional evaluations of the likelihood function.
\newline
Several simulations are provided and the application to a sophisticated astronomical model has been considered, where the number of planets in the system is detected by the analysis of the marginal likelihood. The results show the benefits of the proposed scheme. For instance, in the astronomical example,  the percentage of correct detection of the true model obtained by ATAIS is $\approx 98\%$, whereas with the standard AIS scheme is only $\approx 56\%$.  As future research, we plan to extend the ATAIS scheme in order to deal with an observation model with correlated noise perturbations (for instance, using a Gaussian Process). { Moreover, the use of parallel AIS schemes (or MCMC algorithms) will be also considered. A combination of parallel MCMC chains and AIS schemes can be found in the so-called layered AIS method and other similar approaches  \cite{LAIS17,Botev13}. This idea seems particularly interesting for improving the inference with radial velocity models.   
}

\bibliographystyle{unsrt}
\bibliography{bibliografia}

\appendix 

\section{On the optimization of the likelihood function} 
\label{AppOptimi}
Let us set $\delta=\sigma^2$ and consider to optimize of the  likelihood function
$$
\ell({\bm \theta},\delta) = \frac{1}{(2\pi\delta)^{K/2}} \exp\left(-\frac{V({\bm \theta})}{\delta}\right).
$$
Recall that, in our model, we have $V({\bm \theta})=||{\bf y}-{\bf f}({\bm \theta})||^2$. We desire to obtain
$$
[{\bm \theta}_{\texttt{ML}},\delta_{\texttt{ML}}]=\arg\max \ell({\bm \theta},\delta).
$$
We can write the gradient and equal to zero, 
\begin{gather}
\left\{
\begin{split}
\nabla_{\theta} \ell({\bm \theta},\delta)&= -\frac{1}{\delta} \nabla_{\theta} V({\bm \theta})\left[\frac{1}{(2\pi\delta)^{K/2}}  \exp\left(-\frac{V({\bm \theta})}{\delta}\right)\right]={\bf 0} \Longrightarrow \nabla_{\theta} V({\bm \theta})={\bf 0}, \\
\dfrac{\partial \ell({\bm \theta},\delta)}{\partial \delta}&=\frac{\mathrm{e}^{-\frac{V({\bm \theta})}{\delta}}\,\left(2 V({\bm \theta})-\delta K\right)}{2^{\frac{K}{2}+1} \delta^{\frac{K}{2}+2}\,\pi^{K/2}}=0 
\Longrightarrow \delta=\frac{2}{K}V({\bm \theta}).
\end{split}
\right.
\end{gather}
We have obtained that the ML solution is defined by the system of equations,
\begin{gather}
\left\{
\begin{split}
 &\nabla_{\theta} V({\bm \theta}_{\texttt{ML}} )={\bf 0} \\
 &\delta_{\texttt{ML}}=\frac{2}{K}V({\bm \theta}_{\texttt{ML}} ).
\end{split}
\right.
\end{gather}




\end{document}